\UseRawInputEncoding

\documentclass[review]{elsarticle}

\usepackage{lineno,hyperref}
\modulolinenumbers[5]

\usepackage{graphicx}
\usepackage{subfigure}
\usepackage{epstopdf}
\usepackage{amsmath}
\usepackage{multirow}
\usepackage{color}
\usepackage{amsmath}
\usepackage{amssymb}
\usepackage{mathtools}
\usepackage{mdwmath}
\usepackage{mdwtab}

\begin{document}

\begin{frontmatter}

\title{Direct Symbol Decoding using GA-SVM in Chaotic Baseband Wireless Communication System}


\author[mymainaddress]{Hui-Ping Yin\fnref{myfootnote}}

\author[mymainaddress]{Hai-Peng Ren\corref{mycorrespondingauthor}}
\cortext[mycorrespondingauthor]{Corresponding author}
\ead{renhaipeng@xaut.edu.cn}

\fntext[myfootnote]{Hui-Ping Yin (e-mail:huipingyin@qq.com)}

\address[mymainaddress]{Shaanxi Key Laboratory
of Complex System Control and Intelligent Information Processing, Xi'an University of Technology, Xi'an, China}

\begin{abstract}
To retrieve the information from the serious distorted received signal is the key challenge of communication signal processing. The chaotic baseband communication promises theoretically to eliminate the inter-symbol interference (ISI), however, it needs complicated calculation, if it is not impossible. In this paper, a genetic algorithm support vector machine (GA-SVM) based symbol detection method is proposed for chaotic baseband wireless communication system (CBWCS), by this way, treating the problem from a different viewpoint, the symbol decoding process is converted to be a binary classification through GA-SVM model. A trained GA-SVM model is used to decode the symbols directly at the receiver, so as to improve the bit error rate (BER) performance of the CBWCS and simplify the symbol detection process by removing the channel identification and the threshold calculation process as compared to that using the calculated threshold to decode symbol in the traditional methods. The simulation results show that the proposed method has better BER performance in both the static and time-varying wireless channels. The experimental results, based on the wireless open-access research platform, indicate that the BER of the proposed GA-SVM based symbol detection approach is superior to the other counterparts under a practical wireless multipath channel.
\end{abstract}

\end{frontmatter}


\section{Introduction}

Filter design and coherent/noncoherent decoding are the key points in communication signal processing \cite{Sklar2006Digital}. Researchers from communication field like the analytical decoding methodologies with rigorous theoretical base. However, that might lead to more and more complicated algorithm in complex channel communication \cite{yao2017chaos,ren2019performance,yin2019esn}. By turning around, we treat the binary information decoding in the communication systems as a pattern recognition problem. It can be simply dealt with using the machine learning (ML) approach, which requires no direct analytical support, but works well due to its deepen insight treatment of the signal. In this paper, we illustrate this point using the symbol decoding problem in chaotic baseband wireless communication system \cite{yao2019experimental} as a paradigm.

Owing to its intrinsical properties, such as high sensitivity upon initial conditions, wideband, orthogonality, easy to generate, chaos has gained momentum in providing effective solutions to solve significant practical problems in the communication fields \cite{williams2001chaotic,bollt2003review,chen2017design,9098915}. Since early 1990s, chaos has been used in communication field \cite{hayes1993communicating,rosa1997noise,corron1997new}, mainly focusing on the theoretical analysis in ideal channel with white noise in most of the early research. Until Ref \cite{argyris2005chaos} reported that chaos was successfully used in a commercial wired fiber-optic channel and achieved higher bit transmission rate in 2005, the research on chaos-based communication has turned to practical communication channel. Due to limited bandwidth, multipath propagation, doppler shift, and complex noise, the wireless channel seriously distorts the signal transmitted in it. In terms of this issue, Ref \cite{ren2013wireless} proved the Lyapunov spectrum invariance property of chaotic signal for the first time after transmitting through the wireless channel, which shows chaos can be used to transmit information through wireless channel without information loss. At the same time, more properties of chaos benefiting communication applications have been found, the properties include that chaotic signal has a very simple corresponding matched filter to maximize the signal to noise ratio \cite{corron2010matched,corron2015chaos}, and the chaos property can be used to theoretically remove inter-symbol interference (ISI) caused by multipath propagation completely \cite{yao2017chaos,yao2019experimental}. However, this theoretical ISI free decoding method needs three parts of information including 1) channel information, which can be identified using the probe signal; 2) the past decoded information bits, which are available at the current time for the threshold calculation; 3) the future information bits to be transmitted, which are not available at the current time. To avoid using the future bits, Ref \cite{yao2017chaos} proposed a suboptimal decoding threshold, experimental tests in \cite{yao2019experimental} showed the CBWCS achieved better BER performance using the suboptimal threshold with lower computation cost as compared to the conventional wireless communication system with complex channel equalization. Although the results in \cite{yao2019experimental} are attracting enough, it leaves margin for further improvement. The first attempt was predicting the future chaotic baseband waveform in \cite{ren2019performance}, which was used to decode future information bits. Due to the long term prediction difficulty of the chaotic signal, only one future bit is derived and used together with the past information bits to calculate more accurate threshold. Both simulation and experimental results verified the improved BER performance using the predicted future bit information. The methods in both \cite{ren2019performance} and \cite{yao2019experimental} needed to identify the channel information for decoding. To avoid this operation, the threshold was predicted directly in \cite{yin2019esn} using the echo state network (ESN). This method not only avoids using the identified channel information in decoding but also avoids the error caused by the inaccurate identified channel information, therefore, simulation and experimental results are not surprising to show the superior BER performance with respect to the methods in \cite{ren2019performance} and \cite{yao2019experimental}. All above attempts to improve the BER performance is started from the conventional communication engineering thought, that is, to decode information using single feature, decoding threshold. By considering this problem from pattern recognition viewpoint, the decoding process is a kind of two-category classification problem. Therefore, machine learning (ML) becomes a powerful tool to solve such problems. Following this novel insight, a genetic algorithm (GA) optimized support vector machine (SVM) is proposed to decode the information bit directly in CBWCS in this paper.

As a promising classification method, machine learning has been attracting much research interest from various fields due to numerous successful applications to solve practical problems \cite{er2009theory,xu2016personalized,chen2018flight,hosseinyalamdary2018deep,woo2018unsupervised,yang2018convolutional,coutinho2018learning}. In the field of communication and signal processing, different types of machine learning methods, including unsupervised learning, reinforcement learning, and supervised learning, find significant applications in many cases \cite{jiang2016machine,nguyen2008survey,bi2015wireless}. SVM is one of the machine learning methods that has state-of-the-art capability for classification, and it has been applied in various fields \cite{guenther2016support,ghaddar2018high,ma2014support}. The idea behind the SVM is to map the input vectors into a high-dimensional feature space in which they become linearly separable. The mapping from the input vector space to the feature space is non-linear which can be implemented in a convenient and efficient way through the kernel functions (i.e., the so-called kernel trick). A SVM classifier is in principle a binary classifier, this means that it can predict or classify input data belonging to two distinct classes. SVM has been employed to design link adaptation methods for wireless communication networks \cite{daniels2009online,yun2009multiclass,rico2014learning}. There are extensive studies on not only improving the accuracy of SVMs \cite{feng2017scalable}, but also applying SVM to various classification problems such as signal detection in visible light communications \cite{yuan2017svm}. However, its usage in communication systems has not been well explored.

In this paper, a SVM is used into CBWCS to decode the information bit directly according to the waveform received. The SVM is utilized for the classification task \cite{ben2010user,chang2011libsvm}, and its parameters are optimized by the genetic algorithm (GA) to explore the best classification ability of SVM. The global optimizing GA helps to decrease the SVM structure risk and rent the SVM ability to use small training data set in order to achieve desired generalization performance. This point is very important for practical communication system application because larger set of training data means the channel bandwidth is occupied more.

The main contributions of this work are summarized as follows: first, we identify the information symbol decoding process for chaotic baseband wireless communication as a binary classification problem for the first time; second, we suggest an appropriate learning approach, including the training data with the corresponding class labels and GA based SVM parameters optimization; third, we test the proposed method in both simulation and experiment to show its superiority with respect to the comparison methods.

Numerical simulations and experiments based on software defined radio system are extensively performed in order to evaluate the proposed GA-SVM based symbol detection method. The proposed method achieves better performance with simplified decoding process as compared to the decoding methods using the calculated threshold \cite{yao2017chaos,ren2019performance,yao2019experimental}.

The rest of the paper is organized as follows. In Section 2, we introduce the GA-SVM based symbol detection approach in chaotic baseband wireless communication system. In Section 3, the performance of the GA-SVM based symbol detection method is compared with the other symbol detection methods using decoding threshold in simulation. In Section 4, the proposed method is evaluated in experiment based on wireless open-access research platform (WARP) system. Finally, we conclude this paper in Section 5.

\section{GA-SVM based symbol decoding}
\subsection{Working principle of CBWCS}
The block diagram of CBWCS is given in figure 1. From figure 1, we know that the information bits (symbols) to be transmitted are sent to the chaotic shape-forming filter (CSF) \cite{renchaotic} to generate the chaotic baseband signal $x\left( t \right)$, and the baseband signal is fed into the up-carrier for channel modulation, then, the chaotic signal is transmitted through wireless channel. At the receiver, the signal is captured by antenna, and fed into down-carrier to remove the carrier frequency. The down-carrier output $r\left( t \right)$, is fed into the corresponding matched filter (MF) in order to get the maximum signal to noise ratio (SNR) at the predefined sampling point from the filter output signal $y\left( t \right)$. The sampling point with maximum SNR is compared with the calculated threshold to decode the symbol (bit) transmitted.

\begin{figure}[ht]
  \centering
  \includegraphics[width=3.5in,height=2.2in]{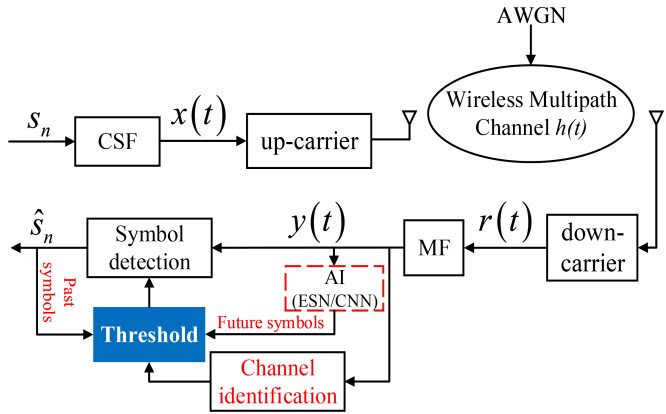}\\
  \caption{Block diagram of the threshold-based symbol detection in CBWCS.}
\end{figure}

The threshold used in the CBWCS is one of the key point from conventional communication engineering viewpoint. Zero could be used as threshold, but it has the worst performance, because this means the inter-symbol interference (ISI) from past and future symbols is not considered \cite{ren2019performance}. Suboptimal threshold is proposed in \cite{yao2017chaos}, which is calculated by using the decoded past symbols and the channel parameters identified from the probe stream \cite{yao2019experimental}. To further consider the future symbol effect, an echo state network (ESN) is proposed in \cite{ren2019performance} to predict the future output waveform of the matched filter, from which the future one symbol is predicted, finally, the decoded one future symbol is used together with the past decoded symbols and the identified channel information in order to obtain more accurate decoding threshold, in this way, the BER performance is improved. In order to avoid the waveform prediction difficulty in \cite{ren2019performance}, reference \cite{yin2019esn} proposes to predict the decoding threshold directly, instead of predicting waveform then decoding a future symbol from the waveform, which simplifies the procedure and achieves better BER performance, because the effect of imperfect waveform prediction is removed. Here, one point to be noticed is that not only ESN could be used for the symbol prediction but also other AI methods, like convolutional neural network (CNN) can also play such a role \cite{ren2020artifical} as shown in figure 1.

Now we conclude the above mentioned methods. First of all, all these methods have the same start point, that is, to calculate the threshold derived in \cite{yao2017chaos}, which is then used to decode symbol. Second, because of this start point, complicated computations are required, including 1) the channel parameters identification, 2) the future waveform prediction and/or 3) the future symbol prediction, and 4) the threshold calculation. All these procedures bring the error contribution to the final BER.

In fact, could we jump out of the start point above, from another viewpoint, we can treat the decoding process as a pattern recognition problem, that is, classification of two categories from the filter output signal. By this way, the complicated channel parameters identification is removed, the complicated future symbol prediction methods are avoided, therefore, the BER contribution of these procedures is eliminated, which promises to decrease BER.

In the following section, we will give the framework of CBWCS based on direct GA-SVM symbol decoding.

\subsection{Symbol decoding based on GA-SVM in CBWCS}
Inspired by the pioneering works based on learning-based method in wireless communication, we focus on the symbol detection in chaotic baseband wireless communication system using machine learning approach. The symbol detection in wireless communication is converted to be a binary classification problem. Compared to the symbol detection method using decoding threshold \cite{yao2017chaos,ren2019performance,yao2019experimental}, the GA-SVM model provides relatively simplified process.

After we change the start point from the decoding threshold to the final decoding task. We design a simplified framework of CBWCS based on GA-SVM symbol decoding as shown in figure 2, where the decoding part is significantly simplified as the block marked by green with respect to the corresponding part in figure 1.

\begin{figure}[ht]
  \centering
  \includegraphics[width=3.8in,height=0.9in]{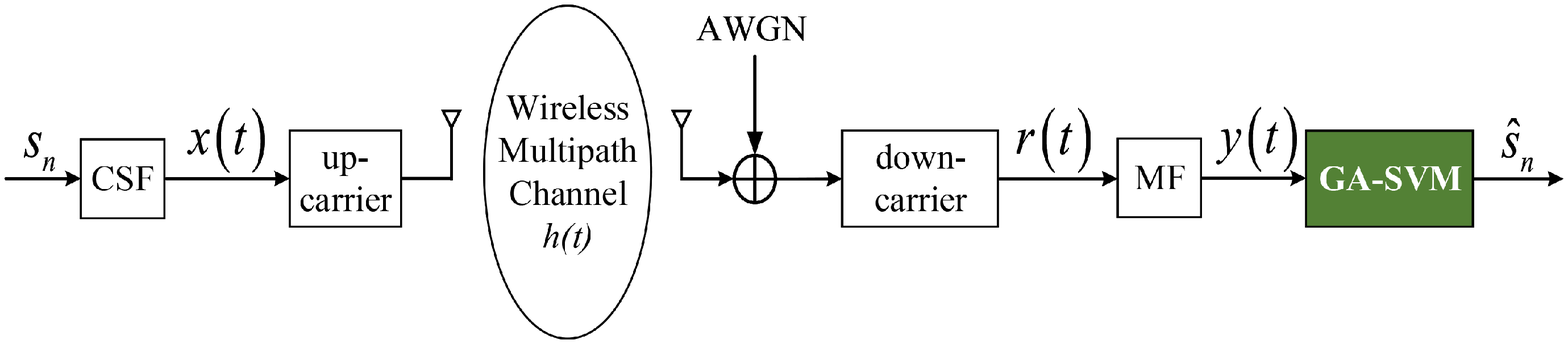}\\%
  \caption{Block diagram of the symbol decoding based on GA-SVM in CBWCS.}
\end{figure}

From figure 2, we know that one piece of matched filter output is used as input of the classifier, and the classifier output is the decoding symbol.

From the pattern recognition viewpoint, machine learning has achieved lots of advancement \cite{bishop2006pattern,wang2009algorithm,Wang2018Feature,li2018hyperspectral,chun2018recognition,bulgarevich2018pattern}. The power of machine learning depends on the training data set and the proper classifier structure. SVM is considered as the most powerful classifier \cite{ghaddar2018high,ruan2019granular,guo2018effective} by employing the high dimensional feature representation provided if the proper structure is selected.

In the following sub-sections, we tackle the problems of training data set and the SVM design with minimum structure risk.

\subsection{Training data set}
To design training data set, we first analyze the feature of the baseband signal in CBWCS \cite{yao2019experimental}. The baseband signal is generated by a chaotic shape-forming filter (CSF) \cite{renchaotic} as shown in figure 2, given by $x\left( t \right) = \sum\nolimits_{n =  - \infty }^\infty  {{s_n} \cdot p\left( {t - {n \mathord{\left/
 {\vphantom {n f}} \right.
 \kern-\nulldelimiterspace} f}} \right)} $, where the basis function $p\left( t \right)$ is given by

\begin{equation}
 p\left( t \right)\! =\! \left\{ \begin{array}{l}
\left( {1\! -\! {e^{ - {\beta  \mathord{\left/
 {\vphantom {\beta  f}} \right.
 \kern-\nulldelimiterspace} f}}}} \right){e^{\beta t}}\left( {\cos \omega t \!- \!\frac{\beta }{\omega }\sin \omega t} \right),\left( {t < 0} \right)\\
1\! - \!{e^{ - \beta \left( {t - {1 \mathord{\left/
 {\vphantom {1 f}} \right.
 \kern-\nulldelimiterspace} f}} \right)}}\left( {\cos \omega t\! -\! \frac{\beta }{\omega }\sin \omega t} \right),\left( {0 \le t < {1 \mathord{\left/
 {\vphantom {1 f}} \right.
 \kern-\nulldelimiterspace} f}} \right)\\
0,\left( {t \ge {1 \mathord{\left/
 {\vphantom {1 f}} \right.
 \kern-\nulldelimiterspace} f}} \right)
\end{array} \right.,
\end{equation}\\
where $\omega $ and $\beta $ are parameters of the basis function satisfying $\omega  = 2\pi f$, $0 < \beta  \le f \cdot \ln 2$, and $f$ is the base frequency.

Figure 3 shows the basis function plot for $\beta  = \ln 2$ and $f = 1$.
\begin{figure}[ht]
  \centering
  \includegraphics[width=3.0in,height=2.0in]{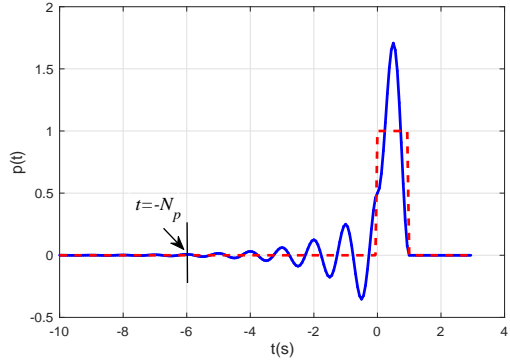}\\
  \caption{The basis function $p\left( t \right)$ for $\beta  = \ln 2$ and $f = 1$.}
\end{figure}

From figure 3, we know that $x\left( t \right)$ can be approximated by
\begin{equation}
x\left( t \right) = \sum\limits_{n = \left\lfloor t \right\rfloor }^{\left\lfloor t \right\rfloor  + {N_p}} {{s_n} \cdot p\left( {t - n} \right)},
\end{equation}
where $\left\lfloor t \right\rfloor $ indicates the largest integer less than or equal to $t$, and ${N_p}$ is a positive integer. From figure 3, we can see that when ${N_p} = 6$, it satisfies $p\left( t \right) \approx 0$ for $t <  - {N_p}$, thus, the value of parameter ${N_p}$ is set to be 6 in this paper. A statistical average model of wireless channel \cite{meinila2009winner} can be defined
\begin{equation}
h\left( t \right) = \sum\limits_{l = 0}^{L - 1} {{\alpha _l}} \delta \left( {t - {\tau _l}} \right),
\end{equation}
where ${\alpha _l}$ and ${\tau _l}$ are the attenuation and propagation delay corresponding to path $l$ from the transmitter to the receiver, and $\delta \left(  \cdot  \right)$ is the Dirac delta function. Assuming the delay ${\tau _l}\left( {l = 0,1, \ldots ,L - 1} \right)$ satisfies $0 = {\tau _0} < {\tau _l} <  \cdots  < {\tau _{L - 1}}$, the channel fading ${\alpha _l}$ can be modeled as a negative exponential decay ${\alpha _l} = {e^{ - \gamma {\tau _l}}}$, where $\gamma $ is the damping coefficient \cite{ren2013wireless}.

 At the receiver of the CBWCS, the received signal is given by
 \begin{equation}
 \begin{aligned}
r\left( t \right) &= h\left( t \right) * x\left( t \right) + w\left( t \right)\\
                  &= \sum\limits_{l = 0}^{L - 1} {{\alpha _l}x\left( {t - {\tau _l}} \right)}  + w\left( t \right)\\
                  &= \sum\limits_{l = 0}^{L - 1} {{\alpha _l}\sum\limits_{n =  - \infty }^\infty  {{s_n}p\left( {t - {\tau _l} - n/f} \right)} }  + w\left( t \right),
\end{aligned}
 \end{equation}
where '$ * $' denotes convolution and $w\left( t \right)$ is an Additive White Gaussian Noise (AWGN).

The filter output is given by
\begin{equation}
\begin{aligned}
y\left( t \right) &= g\left( t \right) * r\left( t \right)\\
 &= \int_{\tau  =  - \infty }^\infty  {p\left( { - \tau } \right)} r\left( {t - \tau } \right)d\tau \\
 &= \sum\limits_{l = 0}^{L - 1} {{\alpha _L}} \sum\limits_{n =  - \infty }^\infty  {{s_n}} \left[ {\int_{\tau  =  - \infty }^\infty  {p\left( \tau  \right)p\left( {\tau  - t + {\tau _l} + \frac{n}{f}} \right)d\tau } } \right]\\
 &+ \int_{\tau  =  - \infty }^\infty  {p\left( {\tau  - t} \right)w\left( \tau  \right)d\tau },
\end{aligned}
\end{equation}
where $g\left( t \right) = p\left( { - t} \right)$ is the impulse response of the matched filter (MF) \cite{corron2010matched}.

From Eqs. (4) and (5), we learn that the received chaotic baseband signal of the current symbol is affected by both the past symbols and the future ones through the multipath wireless channel. This fact makes it possible for us to use the received signal (the past, the current, the future) to decode the current symbol and at the least the past and future ${N_p}$ symbols affect the decoding of the current symbol.

In the following, we analyze the influence of different combinations of past and future symbols on the waveform corresponding to the current symbol as shown in figures 4 and 5.

\begin{figure}[ht]
  \centering
  \includegraphics[width=4.0in,height=2.4in]{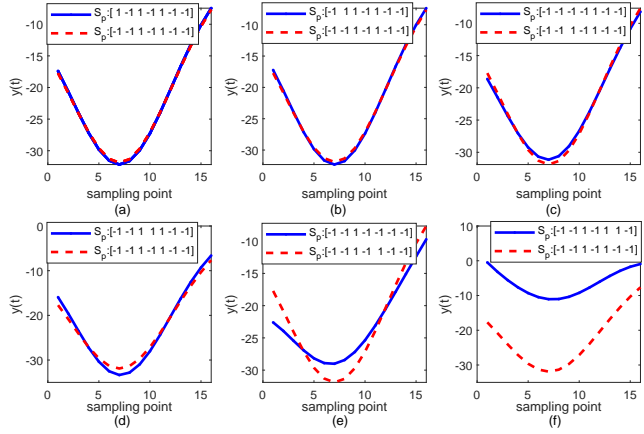}\\
  \caption{The current symbol waveform difference caused by past symbols.}
\end{figure}

\begin{figure}[ht]
  \centering
  \includegraphics[width=4.0in,height=2.4in]{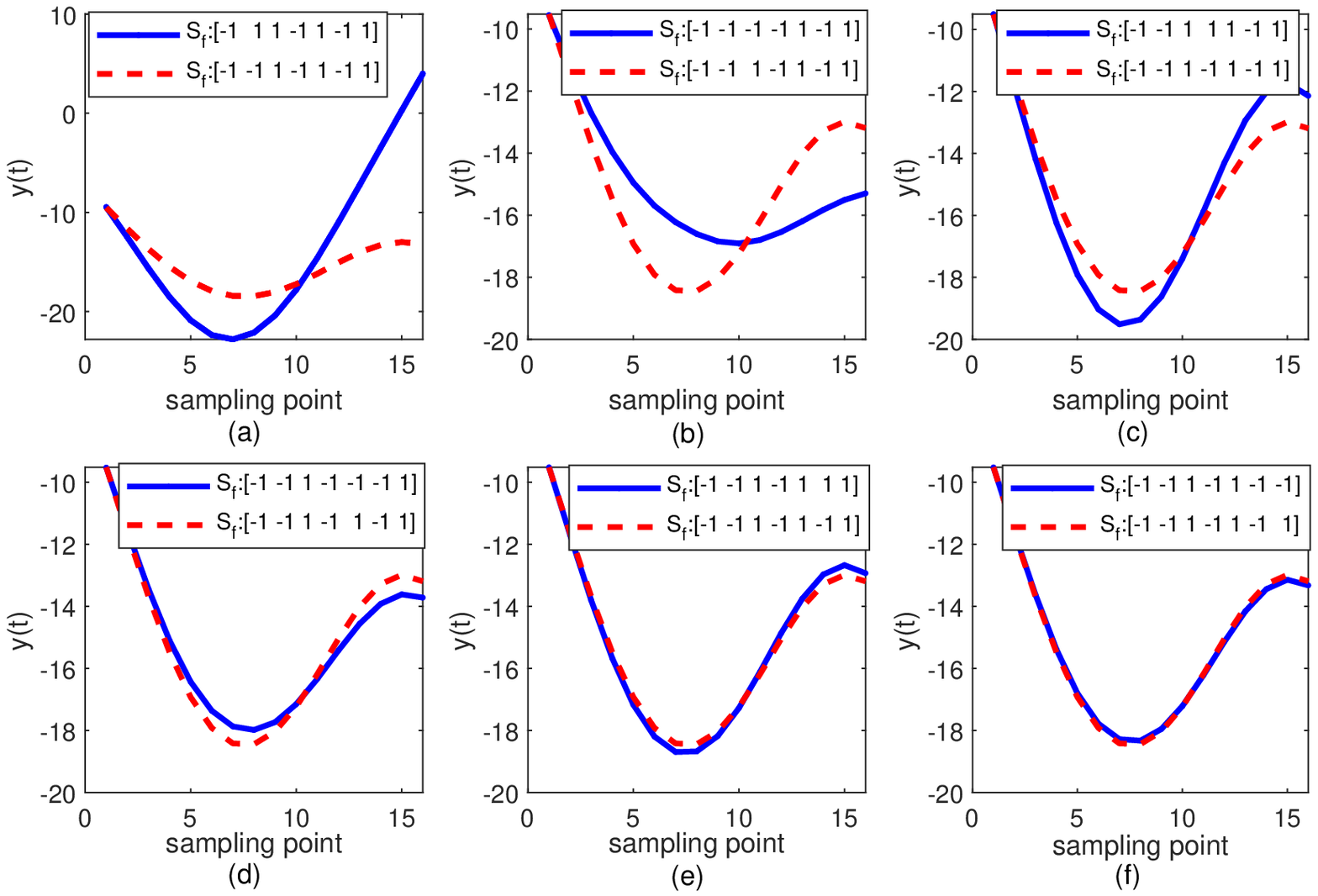}\\
  \caption{The current symbol waveform difference caused by future symbols.}
\end{figure}

We assume the current symbol ${s_n} =  - 1$, and the corresponding chaotic waveforms are represented by the blue solid line and red dashed line respectively with different combinations of past and future symbols, where ${S_p} = \left[ {{s_{n - 6}}, \cdots ,{s_n}} \right]$ represents the past and current symbols set, ${S_f} = \left[ {{s_n}, \cdots ,{s_{n + 6}}} \right]$ represents the current and future symbols set. We can see that only a little difference is caused by different ${s_{n - 6}}$ in figure 4(a) and by different ${s_{n - 5}}$ in figure 4(b); figure 4(c)-(f) show the differences caused by different ${s_{n - 4}}$, ${s_{n - 3}}$, ${s_{n - 2}}$, ${s_{n - 1}}$ respectively; figure 5(a)-(f) show the differences caused by different ${s_{n + 1}}$, ${s_{n + 2}}$, ${s_{n + 3}}$, ${s_{n + 4}}$, ${s_{n + 5}}$, ${s_{n + 6}}$ respectively.

From figures 4 and 5, we know that the waveform of the current symbol is influenced by both the past and future symbols, the closer to the current symbol the bit difference locates, the bigger effect on the waveform corresponding to the current symbol, 4 past and 4 future symbols make main contributions to the current symbol waveform.

To decode the current symbol, the filter output signal corresponding to four past symbols, the current symbol, four future symbols is needed to train the SVM model. 4 past symbols, 1 current symbol, 4 future symbols could generate ${2^9}$ possible combinations, each of which consists of waveform corresponding to 9 bits. So we have the waveform corresponding to 4608 symbols as training data source. But 4608 bits as training data will cost too much time, the efficiency is decreased accordingly. By lessen the restriction, we select the signal corresponding to 3 past symbols, 1 current symbol, and 3 future symbols as the training data set, i.e., 896 bits, thus the amount of training data is greatly reduced. In what follows, we will compare different data sets to see the performance.

In the proposed method, the transmission data is sent frame by frame, the frame structure is shown in figure 6.
\begin{figure}[ht]
  \centering
  \includegraphics[width=2.2in,height=0.3in]{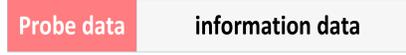}\\
  \caption{The data frame in the wireless communication system.}
\end{figure}

The data frame configuration is compatible with other previous methods and the conventional methods, which also helps to cope with the time variant channel state \cite{yao2019experimental}. one data frame consists of two parts, i.e., the probe data and information data. The probe data is also called training data, which is used for frame synchronization, clock synchronization, and the training data of AI method \cite{ren2019performance,yin2019esn} as well, which is known to the receiver. Whereas, the information data is the data to be transmitted. The frame data including the probe data and information data is passed through the chaotic shape-forming filter to obtain the baseband signal, then, the baseband signal is passed through the up-carrier as shown in figure 2, which is transmitted through the transmission antenna. After the received signal is captured by the receiver antenna, the signal is passed through the down-carrier, and matched filter, then the matched filter output corresponding to a frame is divided into two parts, the probe data part is used for frame synchronization, clock synchronization, and SVM training, it is also used for ESN training in the comparison methods. Afterwards, the information data is decoded using the trained SVM and different comparison algorithms. This point guarantees the fair comparison of the different methods. In this paper, the probe data length can be either 4608 or 896, which will be investigated in the following parts.
\subsection{GA-SVM}

A SVM classifier is in principle a binary classifier. This means that it can predict or classify input data belonging to two distinct classes. SVMs have proved to be a very efficient machine learning technique \cite{ma2014support}.

The idea behind the SVMs is to map the input vectors into a high-dimensional feature space in which the features become linearly separable. The mapping from the input vector space to the feature space is non-linear which can be done in a convenient and efficient way through kernel functions $K\left( {{{\bf{u}}_{i}},{{\bf{u}}_{j}}} \right)$. Given two vectors as ${{{\bf{u}}_{i}}}$ and ${{{\bf{u}}_{j}}}$ from input space, the output result in the dot product of their mapping in the high-dimensional feature space is

\begin{equation}
  K\left( {{{\bf{u}}_{i}},{{\bf{u}}_{j}}} \right) = \left( {\Phi \left( {{{\bf{u}}_{i}}} \right) \cdot \Phi \left( {{{\bf{u}}_{j}}} \right)} \right),
\end{equation}
where $\Phi $ is an explicit mapping from input space to features space \cite{hofmann2008kernel}. Details about the SVM theory can be found in many excellent tutorials and textbooks \cite{guenther2016support,ben2010user}.

Depending on the applications, different types of kernel functions can be used. The radial basis function (RBF) kernel, $K\left( {{{\bf{u}}_i},{{\bf{u}}_j}} \right) = \exp \left( { - g||{{\bf{u}}_i} - {{\bf{u}}_j}|{|^2}} \right)$, is used in our work.

We assume that the training data set is $D = \left\{ {\left( {{{\bf{u}}_1},{v_1}} \right),\left( {{{\bf{u}}_2},{v_2}} \right), \cdots ,\left( {{{\bf{u}}_m},{v_m}} \right)} \right\}$, where ${\bf{u}}$ indicates the input vector, the corresponding class label is ${v} \in \left\{ { - 1, + 1} \right\}$, and sub-index $m$ is the number of training data. Two classes can be separated by a hyperplane ${{\bf{W}}^T}\Phi \left( {\bf{u}} \right) + b = 0$, and it can be converted to be a constrained minimization problem given by
\begin{equation}
\begin{aligned}
&\mathop {\min }\limits_{{\bf{W}},b} \quad \frac{1}{2}||{\bf{W}}|{|^2} + C\sum\limits_{i = 1}^m {{\xi _i}}\\
& \begin{array}{r@{\quad}r@{}l@{\quad}l}
s.t.&{v_i}\left( {{{\bf{W}}^T}\Phi \left( {{{\bf{u}}_i}} \right) + b} \right) \ge 1 - {\xi _i}, i=1,2,3,\ldots,m\\
\end{array},
\end{aligned}
\end{equation}
where $C$ is the penalty term, $\xi$ is a slack variable.

The penalty parameter $C$ and the kernel function parameter $g$ affect the classification accuracy of SVM, in most cases, they are chosen by experience. In this paper, the genetic algorithm (GA) is used to search for better combinations of the parameters. It can obtain the optimal solution after a series of iterative computations \cite{huang2006ga} using the training data. After the GA optimization, a best combination of the parameters fit for the training data is obtained, we refer it to as GA-SVM in this paper.

The optimal values for ${\bf{W}}$ and $b$ can be found by solving the constrained minimization problem in Eq. (7) using the Lagrangian multiplier ${\alpha _i}\left( {i = 1, \cdots ,m} \right)$ \cite{cristianini2000introduction}. Thus, the optimal discriminant function is given by
\begin{equation}
\begin{aligned}
f\left( {\bf{u}} \right) = {\rm{sgn}}\left( {\sum\limits_{i = 1}^m {{\alpha _i}{v_i}K\left( {{{\bf{u}}_i},{\bf{u}}} \right) + b} } \right).
\end{aligned}
\end{equation}

In the training data, those ${{{\bf{u}}_i}}$ with nonzero ${\alpha _i}$ are the ``support vectors'', if we assume that ${N_s}$ is the number of the support vectors, Eq. (8) can be rewritten by

\begin{equation}
\begin{aligned}
f\left( {\bf{u}} \right) = {\rm{sgn}}\left( {\sum\limits_{s = 1}^{{N_s}} {{\alpha _s}{v_s}K\left( {{{\bf{u}}_s},{\bf{u}}} \right) + b} } \right),
\end{aligned}
\end{equation}
where ${{\bf{u}}_s}$ denotes the support vector, ${{\alpha _s}}$ is the nonzero value of the corresponding Lagrangian multiplier.

The block diagram of GA-SVM can be shown by figure 7.
\begin{figure}[ht]
  \centering
  \includegraphics[width=3.0in,height=2.0in]{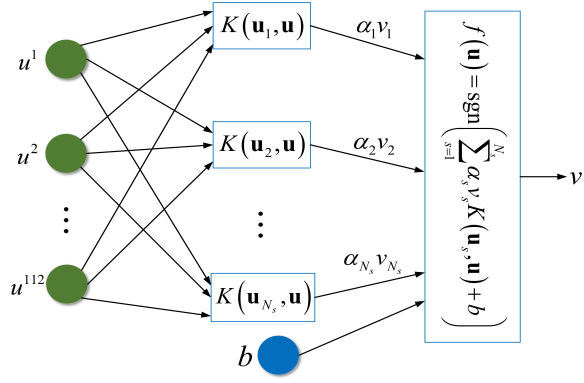}\\
  \caption{The block diagram of GA-SVM.}
\end{figure}

As shown in figure 7, the input vector of the model is ${\bf{u}} = {\left( {{u^1},{u^2}, \cdots ,{u^{112}}} \right)^T}$, where ${u^1},{u^2}, \cdots ,{u^{112}}$ are the matched filter output signal samples corresponding to 3 past symbols, one current symbol and 3 future symbols, where the over-sampling rate ${N_r} = 16$, and the output of the model, $v$, is equal to the current symbol information, i.e., the class label.

Then the trained model using the probe data can be used to decode the symbol from the information data waveform in the CBWCS. For decoding the symbol ${s_n}$, the input vector, ${{\bf{u}}_{n}}$, is fed into the trained GA-SVM model, the corresponding output of the model is the current symbol information, i.e., ${{\hat s}_n} = f\left( {{{\bf{u}}_n}} \right)$, which is shown as figure 8.

\begin{figure}[ht]
  \centering
  \includegraphics[width=3.5in,height=2.6in]{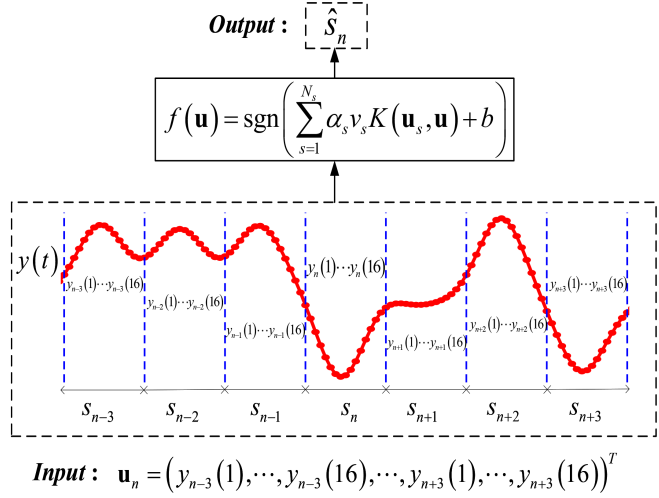}\\
  \caption{The symbol decoding schematic diagram using GA-SVM.}
\end{figure}

\section{Simulation results}
In this section, we present the simulation results for the GA-SVM based symbol decoding in the chaotic baseband wireless communication system.

To eliminate the effect of inter-symbol interference (ISI), an optimal decoding threshold ${\theta _n}$ is used for detecting the symbol ${s_n}$ \cite{yao2017chaos}, which can be given by

\begin{equation}
\begin{aligned}
{\theta _n} = I = {I_{past}} + {I_{future}},
\end{aligned}
\end{equation}
where ${I_{past}} = \sum\limits_{l = 0}^{L - 1} {\sum\limits_{i =  - \infty }^{ - 1} {{s_{n + i}}{I_{l,i}}} } $, ${I_{future}} = \sum\limits_{l = 0}^{L - 1} {\sum\limits_{i = 1}^\infty  {{s_{n + i}}{I_{l,i}}} } $, and

\noindent${I_{l,i}} = {\alpha _l}\int_{\tau  =  - \infty }^\infty  {p\left( \tau  \right)} p\left( {\tau  + {\tau _l} + \frac{i}{f}} \right)d\tau $ can be calculated by

\begin{equation}
\begin{aligned}
{I_{l,i}} = \left\{ {\begin{array}{*{20}{l}}
{\begin{array}{*{20}{l}}
{{\alpha _l}\left( {C - {D^{ - 1}}} \right)\left( {A\cos \left( {\omega {\tau _l}} \right) + B\sin \left( {\omega {\tau _l}} \right)} \right),}\\
{if\left( {|{\tau _l} + \frac{i}{f}| \ge \frac{1}{f}} \right)}
\end{array}}\\
{\begin{array}{*{20}{l}}
{{\alpha _l}\left\{ {A\left( {C - D} \right)\cos \left( {\omega {\tau _l}} \right) + B\left( {C + D} \right)\sin \left( {\omega {\tau _l}} \right) + 1 - |{\tau _l}f + i|} \right\},}\\
{if\left( {0 \le |{\tau _l} + \frac{i}{f}| < \frac{1}{f}} \right),}
\end{array}}
\end{array}} \right.
\end{aligned}
\end{equation}
where $A = \frac{{\left( {{\omega ^2} - 3{\beta ^2}} \right)f}}{{4\beta \left( {{\omega ^2} + {\beta ^2}} \right)}}$, $B = \frac{{\left( {3{\omega ^2} - {\beta ^2}} \right)f}}{{4\omega \left( {{\omega ^2} + {\beta ^2}} \right)}}$, and $C = {e^{ - \beta |{\tau _l} + \frac{i}{f}|}}\left( {2 - {e^{ - \frac{\beta }{f}}}} \right)$, $D = {e^{\beta |{\tau _l} + \frac{i}{f}|}}{e^{ - \frac{\beta }{f}}}$.

The BER using the optimal threshold ${\theta _n} = I$ can be calculated by

\begin{equation}
\begin{aligned}
p\left( {error|{\theta _n} = I} \right) = \frac{1}{2}erfc\left( {\frac{P}{{\sqrt {2\sigma _w^2} }}} \right),
\end{aligned}
\end{equation}where $erfc( \cdot )$ is the complementary error function, ${{{P^2}} \mathord{\left/
 {\vphantom {{{P^2}} {\left( {2\sigma _w^2} \right)}}} \right.
 \kern-\nulldelimiterspace} {\left( {2\sigma _w^2} \right)}}$ is the filtered signal to noise ratio (SNR), and $P = \sum\nolimits_{l = 0}^{L - 1} {{I_{l,0}}}$ is the sum of the multipath power for the corresponding symbol, $L$ is the number of multipaths, $\sigma _w^2$ is the variance of the Gaussian noise $w$.

The theory bit error rate curves in the simulation results are calculated using Eq. (12) with the corresponding multipath parameters.

\subsection{The BER performance of different training data sets}
The GA-SVM is trained using 896 bits, 4608 bits, corresponding to the combination of 7 bits and 9 bits, separately, as aforementioned.
\begin{figure}[ht]
  \centering
  \includegraphics[width=3.3in,height=2.5in]{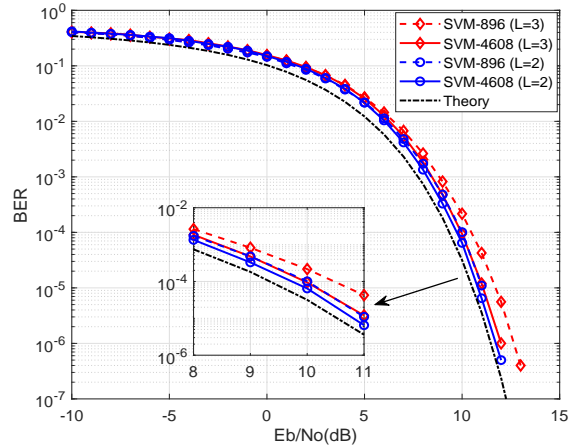}\\
 \caption{The BER results using training data sets with different size under different multipaths. The solid lines represent the BER results using 4608 bits, the dashed lines represent the BER results using 896 bits. The blue lines with circle markers are the BER curves under 2 paths, and the red lines with diamond markers are the BER curves under 3 paths. Black dash-dotted line is the theory BER corresponding to $L = 2$.}
\end{figure}

In figure 9, we investigate the impact of training data sets of different size on the BER of CBWCS by using 896 bits, 4608 bits versus different SNR values. It can be seen that as the number of the training data increases, the BER of the GA-SVM based symbol decoder decreases.

The training data set will grow exponentially considering one more past symbol and one more future symbol correspondingly. The BER comparison results in figure 9 show that the BER performance using 4608 training bits is not improved too much than the performance using 896 training bits, but the number of the training bits is increased greatly, the training complexity is increased correspondingly. By comprehensively considering the performance and cost, 896 bits are used in this paper.

\subsection{BER performance of different methods}
The simulations under different multipath cases using different decoding methods are performed. Figures 10 and 11 show the BER performance comparison for different decoding methods under 2 and 3 paths, respectively. In figures 10 and 11, the $\gamma $ is set to be 0.6, and the time delay parameters for figure 10 are ${\tau _0} = 0$, ${\tau _1} = 1$, for figure 11, ${\tau _0} = 0$, ${\tau _1} = 1$, ${\tau _2} = 2$.
\begin{figure}[ht]
  \centering
  \includegraphics[width=3.3in,height=2.5in]{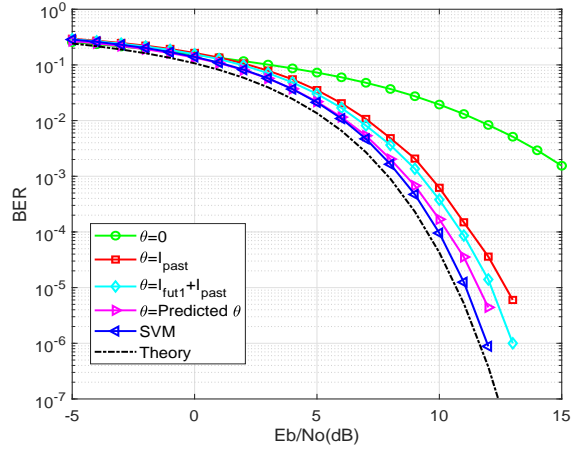}\\
\caption{BER results using different decoding methods under 2 paths. Black dash-dotted line is the theory BER corresponding to $L = 2$, blue solid line with left triangle markers is the BER curve using GA-SVM, magenta solid line with right triangle markers is the BER curve using the decoding threshold predicted by ESN, cyan solid line with diamond markers is the BER curve using the decoding threshold $\theta  = {I_{fut1}} + {I_{past}}$, red solid line with square markers is the BER curve using the decoding threshold $\theta  = {I_{past}}$, green solid line with circle markers is BER curve using zero threshold.}
\end{figure}

\begin{figure}[ht]
  \centering
  \includegraphics[width=3.3in,height=2.5in]{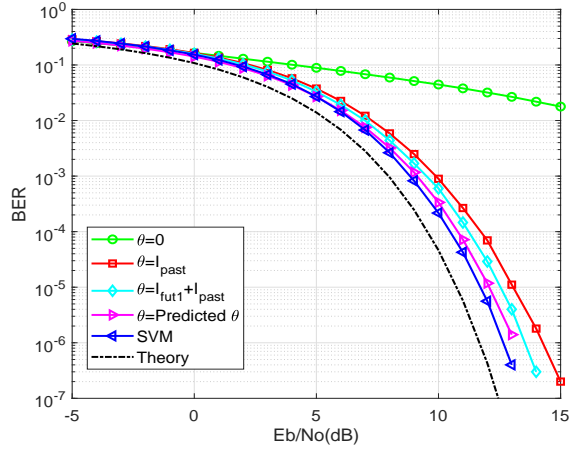}\\
  \caption{ BER results using different decoding methods under 3 paths. Black dash-dotted line is the theory BER corresponding to $L = 3$, blue solid line with left triangle markers is the BER curve using GA-SVM, magenta solid line with right triangle markers is the BER curve using the decoding threshold predicted by ESN, cyan solid line with diamond markers is the BER curve using the decoding threshold $\theta  = {I_{fut1}} + {I_{past}}$, red solid line with square markers is the BER curve using the decoding threshold $\theta  = {I_{past}}$, green solid line with circle markers is BER curve using zero threshold.}
\end{figure}

The results using different decoding methods for two-path channel ($L = 2$) and three-path channel ($L = 3$) in figures 10 and 11 show that the CBWCS has the worst BER by using the decoding threshold without considering the inter-symbol interference (ISI) from past and future symbols, i.e., $\theta  = 0$. Using the decoding threshold $\theta  = {I_{past}} $ in \cite{yao2017chaos} which considered the ISI caused only by past symbols, the BER is lower. Using the decoding threshold $\theta  = {I_{past}} + {I_{fut1}}$ which considered the ISI caused by the past symbols and one future symbol predicted by echo state network (ESN) in \cite{ren2019performance}, the BER is much lower. Using the decoding threshold predicted directly by ESN in \cite{yin2019esn}, the BER performance is further improved as compared to the other methods above. Using the direct decoding based on GA-SVM proposed in this paper, the BER is the lowest among them. With the number of multipath increasing, the distance from theoretical BER to BER of the proposed method increases. But the proposed method is still the best among all methods.

In the simulation cases in figures 10 and 11, the channel parameters are fixed. In fact, the practical channel parameters are time variant. We assume that the channel parameters are unchanged within one data frame, but be variant from one frame to the next. The channel parameter, $\gamma $ obeys the uniform distribution in the range of [0.3, 0.9]. Time delay are ${\tau _0} = 0$, ${\tau _1} = 1$ for two-path channel, and ${\tau _0} = 0$, ${\tau _1} = 1$, ${\tau _2} = 2$ for three-path channel. Figures 12 and 13 give the BER comparison results under such quasi-time-varying channel corresponding to $L = 2$ and $L = 3$, separately.
\begin{figure}[ht]
  \centering
  \includegraphics[width=3.3in,height=2.5in]{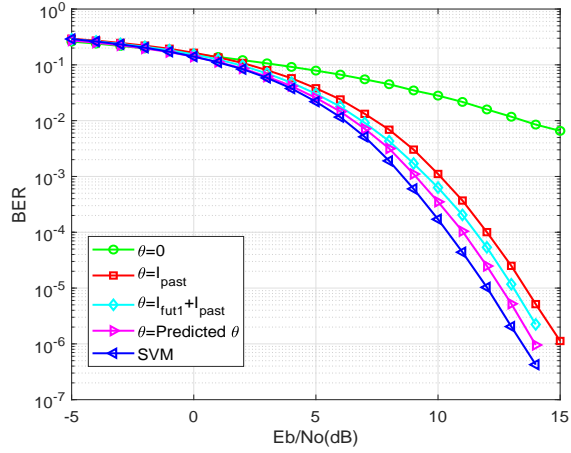}\\
  \caption{ BER results with different decoding methods under 2 paths in a time-varying wireless channel. Blue solid line with left triangle markers is the BER curve using GA-SVM, magenta solid line with right triangle markers is the BER curve using the decoding threshold predicted by ESN, cyan solid line with diamond markers is the BER curve using the decoding threshold $\theta  = {I_{fut1}} + {I_{past}}$, red solid line with square markers is the BER curve using the decoding threshold $\theta  = {I_{past}}$, green solid line with circle markers is BER curve using zero threshold.}
\end{figure}

\begin{figure}[ht]
  \centering
  \includegraphics[width=3.3in,height=2.5in]{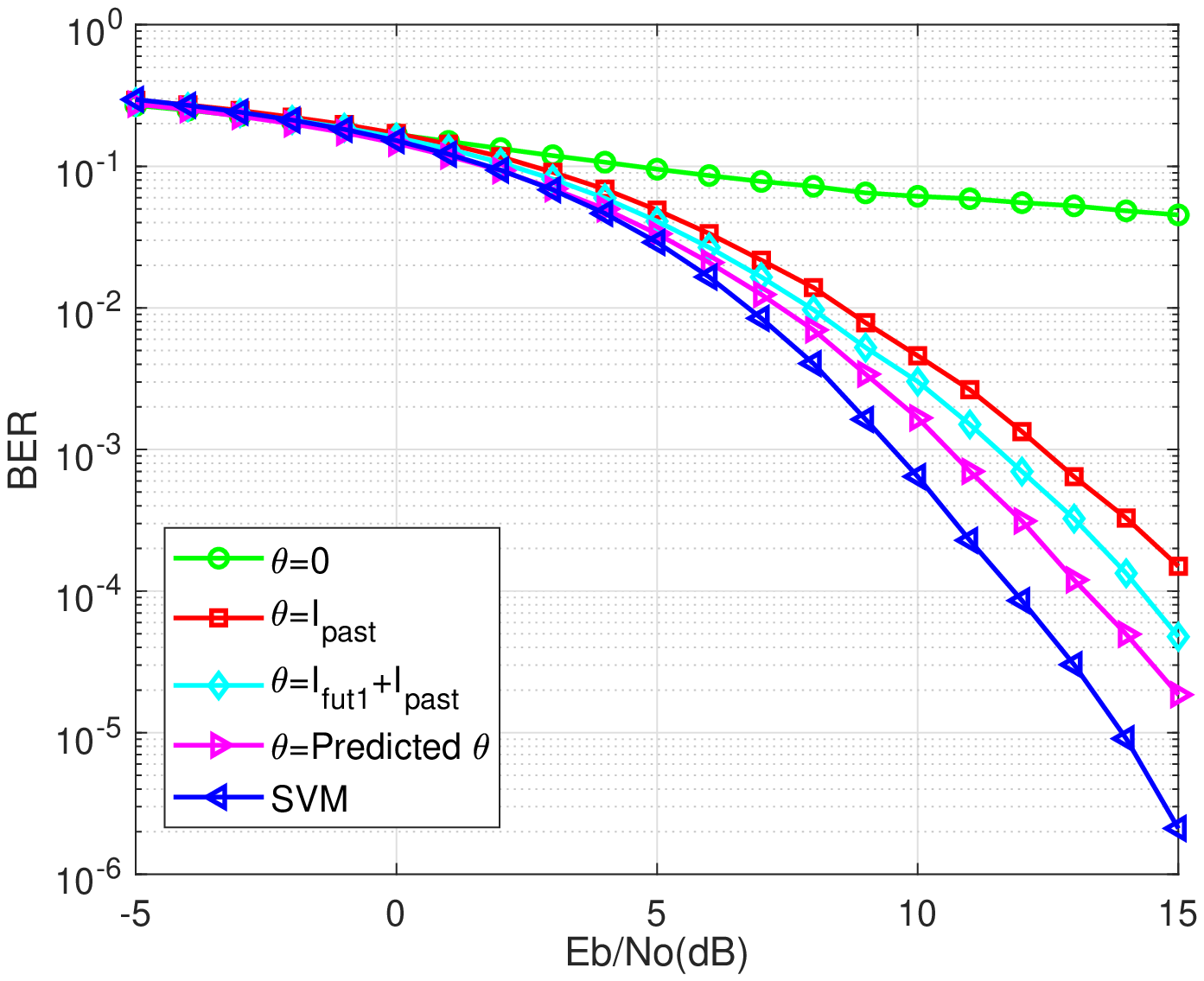}\\
  \caption{ BER results with different decoding methods under 3 paths in a time-varying wireless channel. Blue solid line with left triangle markers is the BER curve using GA-SVM, magenta solid line with right triangle markers is the BER curve using the decoding threshold predicted by ESN, cyan solid line with diamond markers is the BER curve using the decoding threshold $\theta  = {I_{fut1}} + {I_{past}}$, red solid line with square markers is the BER curve using the decoding threshold $\theta  = {I_{past}}$, green solid line with circle markers is BER curve using zero threshold.}
\end{figure}

In our time-varying channel simulation, the frame structure is 896 probe data and 1152 information bits. The probe bits are used for GA-SVM model training, and for the other methods, the probe bits are used not only for AI training, but also for channel parameters estimation.

From figures 12 and 13, the comparison results using the proposed method and other methods show that the proposed method has even better result as compared to the static channel case. The conclusion is the same as that can be drawn from figures 10 and 11, that is, the proposed method has the best performance under the time variant channel.

\section{Experimental results}
To test the proposed method, the experiment is carried out using wireless open-access research platform version 3 (WARP V3) designed by Rice University. For digital signal processing, the Xilinx Virtex6 LX240T FPGA is used, dual-channel and 2.4GHz/5GHz dual-band transceiver are supported by two MAX2829 RF chips, the maximum transmission power is 20dBm by using the dual-band power amplifier, the 12-bit low power analog/digital converter AD9963 is used to provide two ADC channels with sampling rates of 100 MSPS and two DAC channels with sample rates of 170 MSPS, and two 10/100/1000 Ethernet interfaces (Marvell 88E1121R) are used to realize the high-speed signal exchange with the Personal Computer (PC). The photo of the WARP system is shown in figure 14.
\begin{figure}[ht]
  \centering
  \includegraphics[width=2.5in,height=1.3in]{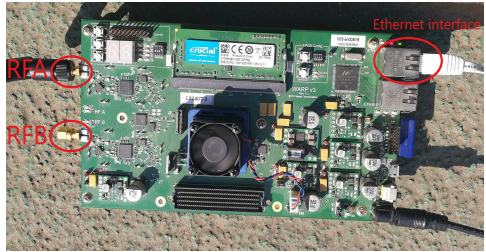}\\
  \caption{The WARP system photo.}
\end{figure}

Each WARP node has two radio antennas, namely RFA and RFB. In our test, only the RFA is used, which is operating at the 5GHz carrier frequency with 20MHz bandwidth.

The BER performance of the GA-SVM based symbol decoding method under the practical channel case is tested. The test is performed at the university campus, and the photo of the test scenario is shown in figure 15, the distance between TX and RX is about 30 meters, there are many trees between them.
\begin{figure}[ht]
  \centering
  \includegraphics[width=2.8in,height=1.8in]{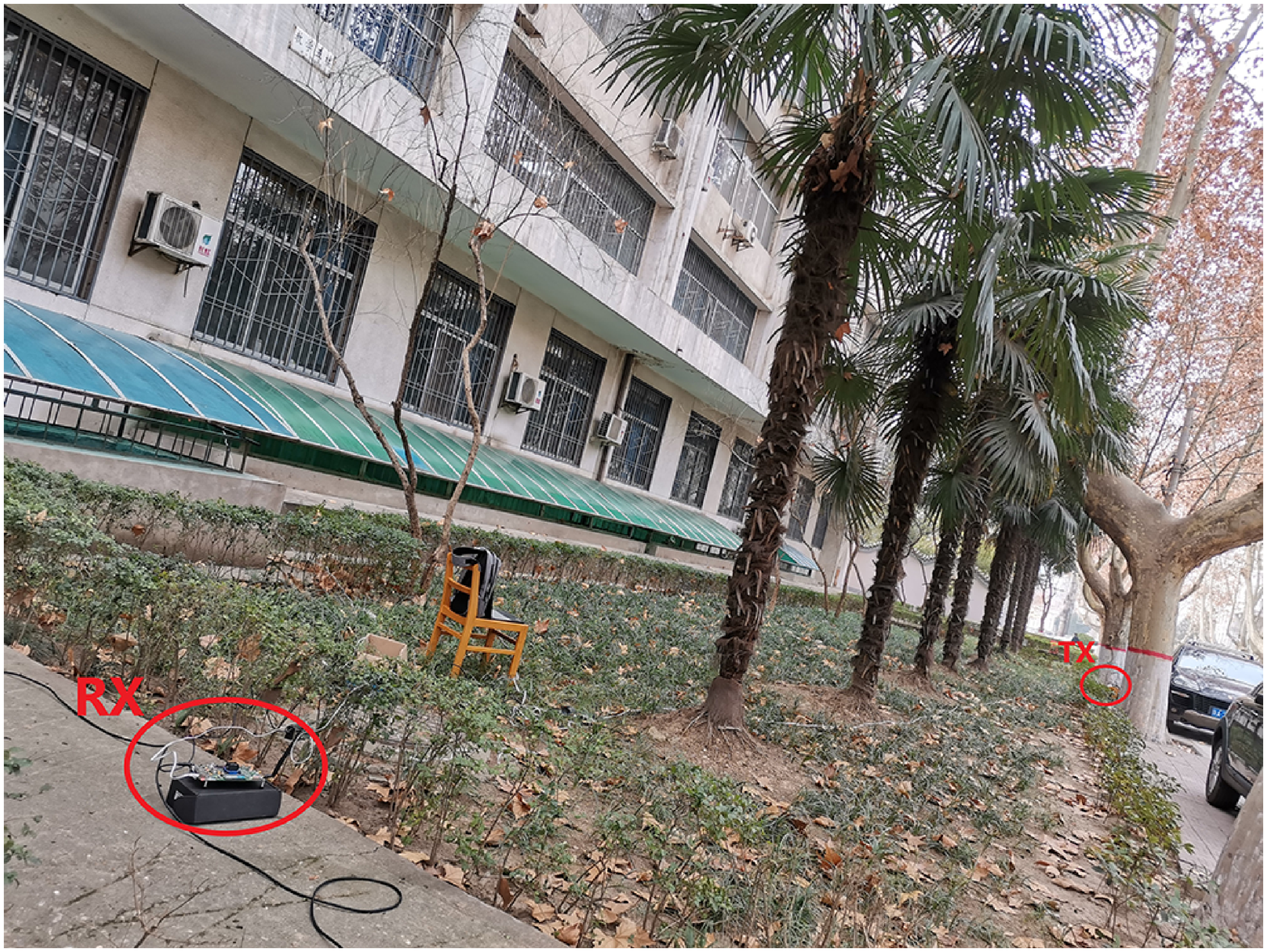}\\
  \caption{The test scenario at the campus.}
\end{figure}

As mentioned before, the information is transmitted frame by frame, there are 2048 bits in one frame, which contains 896 probe data for GA-SVM model training, also used for clock synchronization, 1152 bits are information data.
\begin{figure}[!t]
  \centering
  \includegraphics[width=3.3in,height=2.5in]{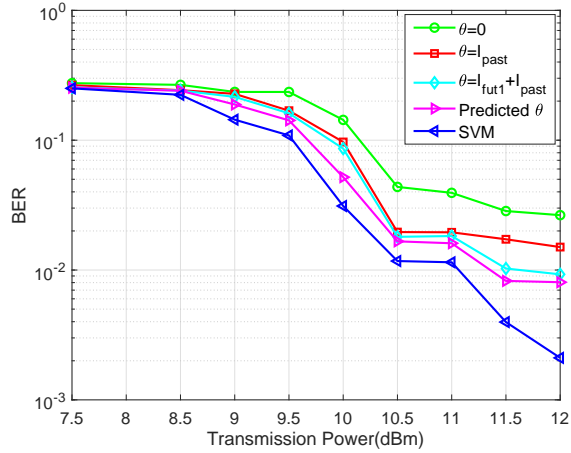}\\
  \caption{The experimental BER curves using different decoding methods. Blue solid line with left triangle markers is the BER curve using GA-SVM, magenta solid line with right triangle markers is the BER curve using the decoding threshold predicted by ESN,cyan solid line with diamond markers is the BER curve using the decoding threshold $\theta  = {I_{fut1}} + {I_{past}}$, red solid line with square markers is the BER curve using the decoding threshold $\theta  = {I_{past}}$, green solid line with circle markers is BER curve using zero threshold.}
\end{figure}

Four different methods are used in the scenario for comparison, the experimental BERs versus transmission power are shown in figure 16, we can see that the method based on GA-SVM get much better performance than the other comparison methods, which is consistent with the simulation results. The channel estimation in practice is so complicated that it is hard to obtain accurate channel state information, so the error contributes to the threshold calculation in these methods using channel information. In the GA-SVM learning process, the effects of the channel identification error and prediction error are excluded, that is one reason why the performance of GA-SVM is the best among the comparison methods. The most important reason to achieve the best performance is that GA-SVM with the minimum structure risk maps the waveform pattern of different symbols and the channel information into the trained GA-SVM structure to the best possible extent, thanks to the power of machine learning.

The point to be noticed here is that the BER curves in figure 16 seem ``abnormal", because, with the transmission power increasing, the BER does not decrease monotonically as intuitively expected. In fact, the underlying assumption of this expectation is that the signal to noise ratio (SNR) is increasing with the increasing of the transmission power, however, practically, it is not such situation, we control the transmission power exactly in experiment, but we cannot guarantee the noise level be the same at different transmission moments. Therefore, in the practical environment, when we increase the transmission power with not very large amount, the SNR might not increase, even in some extreme cases, the SNR is decreased because of the noise level is increased larger relatively. Therefore, we observe the ``abnormal", but very normal BER curves as given in figure 16.
For the same reason, in simulation, we can control the noise level very well (exactly), even we can repeat ``noise". But in practical experiment, we cannot calculate the SNR or Eb/No as done in the simulation, therefore, we cannot compare the experimental results with the simulation results in the same term.
However, this obstacle does not affect the conclusion that we draw from the curves, i.e., under the same SNR (at the same transmission moment), our proposed method is superior to the comparison methods, although the BER for large transmission power might be not lower than that for low transmission power because the SNR might not be increased in this case.

\noindent{\bfseries Discussion:} The real-time implementation of GA-SVM could be accomplished by employing the field programmable gate array (FPGA) resource in the experimental WARP set up. There are also many references \cite{anguita2003digital,pan2013fpga,kumar2017hardware,lopes2019parallel} dealing with the problem, which could be referred. Because the main purpose of this paper is to validate the effectiveness and the superiority of the proposed algorithm as compared to other methods. Hardware implementation is not discussed here any more.

\section{Conclusion}
A novel idea to treat the symbol decoding task in the CBWCS as a pattern classification problem is proposed in this paper. Under the guidance of this idea, GA-SVM is trained using the probe data in the transmission data frame, and then used as a symbol decoder to decode the information. This approach avoids the channel information identification or future symbol prediction as well as the BER increasing due to these processes, by this way, the proposed method simplified the procedure and improved the BER performance. The simulation and experimental results prove the feasibility and superiority of the proposed method. Although the implementation complexity of GA-SVM is considerable, it can be solved using FPGA hardware implementation.

\section*{References}

\bibliography{ref3}

\end{document}